\documentclass[prl,aps,twocolumn,showpacs,amsmath,amssymb,floatfix]{revtex4}

\usepackage[dvips]{graphicx}
\usepackage[normalem]{ulem} 
\DeclareGraphicsExtensions{.eps}

\usepackage[ngerman,english]{babel}
\usepackage{textcomp,stmaryrd}

\newif\ifhyper
\hypertrue

\ifhyper
\usepackage[hypertex]{hyperref}
\hypersetup{%
  a4paper = {true},
  bookmarksnumbered = {true}, 
  bookmarksopen = {true}, 
  bookmarksopenlevel = {2}, 
  breaklinks = {true}, 
  colorlinks = {true}, 
  pdfauthor = {\textcopyright\ Miha\l Karski},
  pdfcreator = {\LaTeX\ with package \flqq hyperref\frqq},
  pdfkeywords = {}
  pdfmenubar = {true},
  pdfstartpage = {1},
  pdfstartview = {FitH}, 
  pdfsubject = {},
  pdftitle = {},
  pdftoolbar = {true},
  plainpages = {false}, 
  urlcolor = {blue} 
}
\fi

\newcommand{\hide}[1]{}

\newif\ifadr
\adrtrue

\begin{document}
\title{Nearest-neighbor detection of atoms in a 1D optical lattice by fluorescence imaging}

\author{M.~Karski}
\email[\emph{Electronic address: }]{karski@uni-bonn.de}
\author{L.~F\"{o}rster}
\author{J.~M.~Choi}
\author{W.~Alt}
\author{A.~Widera}
\author{D.~Meschede}

\affiliation{
  Institut f\"{u}r Angewandte Physik der Universit\"{a}t Bonn,
  Wegelerstrasse 8,
  53115 Bonn,
  Germany}

\date{\today}

\begin{abstract}
We overcome the diffraction limit in fluorescence imaging of neutral atoms in a sparsely filled one-dimensional optical lattice. At a periodicity of $433\,\text{nm}$, we reliably infer the separation of two atoms down to nearest neighbors. We observe light induced losses of atoms occupying the same lattice site, while for atoms in adjacent lattice sites, no losses due to light induced interactions occur. Our method points towards characterization of correlated quantum states in optical lattice systems with filling factors of up to one atom per lattice site.
\end{abstract}


\maketitle

Neutral atoms in optical lattices have been shown to be an ideal system for engineering novel types of strongly correlated quantum states. Quantum correlations between different lattice sites could be induced with Bose-Einstein condensates by precisely adjusting the relevant energy scales through controlling the lattice potential \cite{Gre02, Man03a}.
Detection of these novel states was initially only indirect by observing the collapse and revival of the global matter wave interference pattern in time of flight measurements. In contrast, quantum state tomography, as well as many theoretical proposals to employ these correlations for quantum information processing, require single site detection \cite{SingleSiteDetection}, a technically challenging goal for site separations in the optical wavelength domain.
In a different regime, where potential wells are separated by several micrometers, single atoms could be resolved \cite{Nel07,Sch00}. However, in this regime the relevant energy scales are not well amenable to control via the external potential, therefore the ``standard route'' for the preparation of correlated quantum states sketched above seems to favor site separations in the optical wavelength regime.
 Recently, single site detection has been reported in such a system using focused electron beams from an ultra-high vacuum compatible electron gun \cite{Ger08}. This technique is not easily integrated with many current quantum gas experiments, in which, in contrast, optical imaging by fluorescence light is widely established. The latter has seen great success in other fields, e.g., imaging of single molecules \cite{SingleMolecule}. Comparable success with neutral atoms in optical lattices, however, could not be achieved to date.

In this work, we demonstrate the detection of atom pair separations down to nearest neighbors in a one-dimensional (1D) lattice with optical wavelength periodicity. We overcome the previous restrictions imposed by the diffraction limit \cite{Dot05} with a markedly improved data quality and reduced noise, together with advanced numerical processing of fluorescence images. Such a new degree of precision in detection allows us to directly observe  light induced atom losses and to distinguish between on-site and nearest-neighbor contributions. In contrast, detecting  such loss processes have so far relied on ensemble averages in optical lattice systems \cite{Mir06}, while interacting atoms in a type of atom blockade effect have been investigated in systems where only a single running wave optical trap was present \cite{Sch01}. Our approach, however, combines single atom resolution with single site detection.

In our experiment, we load a small number of Caesium atoms in the periodic dipole potential of a standing-wave laser field -- a 1D optical lattice -- from a magneto-optical trap (MOT). The lattice is formed by two counterpropagating linearly polarized laser beams ($\lambda=865.9\,\text{nm}$) with a waist of $w_{0}=20\,\mu\text{m}$ and a typical power of $P=100\,\text{mW}$ providing a trapping potential with a depth of $U/k_{\text{B}}=0.4\,\text{mK}$, for which atom tunneling is negligible.

\begin{figure}[ht]
  \includegraphics[width=0.90\columnwidth]{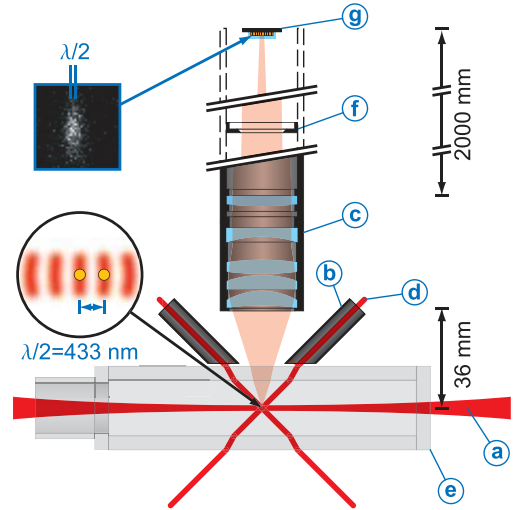}
  \caption{\label{fig:setup}
Detail of the experimental setup. Two counterpropagating laser beams ({\bf a}) form the 1D lattice. Beam tubes ({\bf b}) shield the objective ({\bf c}) from stray light of molasses beams ({\bf d}) off the glass cell ({\bf e}). Apertures ({\bf f}) strongly suppress the remaining stray light. A narrow-band optical filter in front of the EMCCD ({\bf g}) filters the stray light from the optical lattice.
  }
\end{figure}

We illuminate the atoms with a red detuned three-dimensional optical molasses at $852\,\text{nm}$ which also provides continuous Doppler cooling and counteracts heating of the atoms. The fluorescence light is collected by a diffraction limited microscope objective ($\text{NA}=0.29$) \cite{Alt02} and is imaged onto an electron multiplying CCD (EMCCD) camera, where stray light from the molasses beams and the optical lattice are successively filtered, see Fig.~\ref{fig:setup}. For our imaging optics, the diffraction limit in resolution (Airy radius) is given by $1.79\,\mu\text{m}\approx 4\times\lambda/2$. Resolving the lattice periodicity (see later), we infer that one pixel of the EMCCD with a width of $16\,\mu\text{m}$ corresponds to $294.6\,\text{nm}$ in the object plane. 

\begin{figure}[ht]
  \includegraphics[width=0.98\columnwidth]{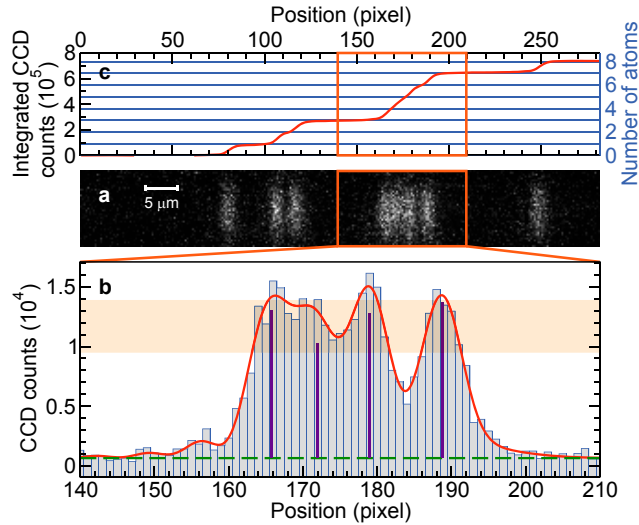}
  \caption{\label{fig:cluster}
({\bf a}) Fluorescence image of atoms. ({\bf b}) Vertically binned intensity distribution $I(x_{i})$ for one region of interest. ({\bf c}) Cumulative sum of $I(x_{i})$ above the background baseline $a_{0}$ (dashed line in {\bf b}). In ({\bf b}), the extracted positions $\xi_{k}$ and the (scaled) fluorescence contributions $a_{k}$ of the atoms are shown (vertical lines). The shaded stripe indicates the acceptance region of the reliability criterion; the solid curve the resulting source distribution $S(x)$ convolved with the LSF.
  }
\end{figure}

The EMCCD image is taken at an exposure time of $1\,\text{s}$. It provides a sampled intensity distribution $\tilde{I}(x_{i},y_{j})$, where $x_{i}$ and $y_{j}$ denote the horizontal and vertical position of pixel $\{i,j\}$, respectively. The intensity distribution of a single atom trapped in the 1D optical lattice shows a characteristic ellipticity, originating from the shape of the trapping potential, see Fig.~\ref{fig:cluster}(a): The atom is vertically confined to the lattice axis by the Gaussian profile of the laser beams, whereas its horizontal position depends on the occupied lattice site. The vertical width of the fluorescence spot is primarily given by the spread of the thermal wave packet of the atom in radial direction of the standing-wave potential. In axial direction, the atoms are strongly confined: The horizontal $1/\sqrt{e}$ half width of the fluorescence spot, corresponding to $\sigma_{\text{sp}}^{\text{hor}}= 810(\pm 19)\,\text{nm}$ in the object plane, is mainly caused by diffraction within the imaging optics (with a theoretical value of $\sigma_{\text{diff}}^{\text{hor}}= 647\,\text{nm}$). Compared with this width, thermal motion of the atoms ($\sigma_{\text{th}}^{\text{hor}}= 23\,\text{nm}$) and drifts of the standing-wave ($12\,\text{nm}/\text{s}$) can be neglected.

To simplify the axial position determination, we bin the intensity distribution vertically $I(x_{i})=\sum_{j}\tilde{I}(x_{i},y_{j})$, see Fig.~\ref{fig:cluster}(b). The resulting distribution is related to the unknown source distribution $S(x)$ by a convolution equation
\begin{equation}\label{eq:convolution}
I(x_{i})=\int_{-\infty}^{\infty}L(x_{i}-u) S(u)\text{d}u+\epsilon(x_{i})\,,
\end{equation}
where $L(x)$ is the area normalized line spread function (LSF) of our imaging optics and $\epsilon(x_{i})$ the additive noise. 

The axial confinement of the atoms ($\sigma_{\text{th}}^{\text{hor}}\ll\sigma_{\text{sp}}^{\text{hor}}$) and systematical suppression of the stray light down to a homogeneous background allow us to model $S(x)$ as
\begin{equation}\label{eq:spike-signal}
	S(x)=a_{0}+\sum_{k=1}^{N}a_{k}\delta(x-\xi_{k})\,,
\end{equation}
where $a_{0}$ denotes the constant baseline of the stray light background, $\delta(x)$ the Dirac delta function representing the strongly confined atom, $a_{k}$ and $\xi_{k}$ the fluorescence contributions and the positions of $N$ atoms, respectively. Therefore, the position determination of the atoms in a 1D lattice corresponds to a parameter estimation of the modeled distribution $S(x)$.

A sufficiently low noise is the first essential prerequisite to deconvolve Eq.~\eqref{eq:convolution} and retrieve the parameters in Eq.~\eqref{eq:spike-signal}. In our experiment, the shot noise of the fluorescence signal dominates, whereas stray light contributions were minimized using light traps (see Fig.~\ref{fig:setup}). Readout noise was reduced by cooling the EMCCD. The remaining noise level is low enough to neglect its signal dependence, which greatly simplifies our numerical method.

The second essential prerequisite is the precise analytical description of the LSF. In principle, the measured intensity distribution of an isolated atom yields information on the LSF, however, a single image does not provide the required resolution and accuracy due to noise and the limited EMCCD resolution. Therefore, we superimpose the intensity distributions of up to hundred isolated atoms with sub-pixel accuracy and precisely fit the shape of the LSF. We restrict the image acquisition to the imaging region with negligible spatial variation of the LSF ($\approx 120\,\mu\text{m}$ in object space).

In order to extract the atomic positions in Eq.~\eqref{eq:spike-signal}, we use a method related to the spike-convolution model fitting~\cite{Li00} which consists of three stages: segmentation, atom number determination and model fitting.\\
In the segmentation, we divide the binned intensity distribution into regions of interest (ROIs) which contain fluorescence from atoms and redundant regions which only contain background. From the latter we estimate the background baseline $a_{0}$. Since parts of our numerical method are based on Fourier transforms, excluding noise from redundant regions allows us to improve both the performance and the precision of our numerical method.

In the following, for each ROI the same procedure is used. We determine the number of atoms $N$ by cumulatively integrating the binned intensity distribution $I(x_{i})$ above the background baseline $a_{0}$. Since each atom contributes equally, the integrated distribution exhibits characteristic steps at integer multiples of the mean single atom fluorescence contribution $I_{\text{a}}$, see Fig.~\ref{fig:cluster}(c). After determining the atom number from $N=\sum_{i\in\text{ROI}}(I(x_{i})-a_{0})/I_{\text{a}}$, we reestimate the (local) baseline $a_{0}$ and determine the atom positions $\xi_{k}$ using the trigonometric moments estimates (for details see Ref.~\cite{Li00}). For each atom, the fluorescence contribution $a_{k}$ is determined by a least squares method, minimizing $\sum_{i\in \text{ROI}}\{I(x_{i})-a_{0}-\sum_{k=1}^{N}a_{k}L(x_{i}-\xi_{k})\}^{2}$. Since in general, trigonometric moments estimates are less precise than maximum likelihood estimates \cite{Li00}, we subsequently improve the parameters $a_{0}$, $a_{k}$ and $\xi_{k}$ using them as an initial guess in a Levenberg-Marquardt fitting algorithm \cite{Lev44}, which then converges within few iterations. For typical images with up to 13 atoms, we determine the atomic positions within less than $100\,\text{ms}$.

The parameters are finally checked for passing a reliability criterion $|a_{k}-I_{\text{a}}|/I_{\text{a}}<20\,\%$ inferred from a statistical analysis of the data \cite{KarskiToBePrep}. By this, we identify and exclude erroneous results, mainly stemming from contributions of atoms which leave the optical lattice during the exposure time \cite{Reliability}.

We apply our numerical method to approx.~6000 images in order to both investigate its efficiency and accuracy, and to extract the distribution of atom separations. For this purpose, we repeatedly load on average $4$ atoms into the optical lattice and successively acquire several images of the same atom distribution. From each image with $N$ atoms we determine the $\tilde{N}\leq N$ atom positions in those ROIs only in which all atoms pass the reliability criterion and calculate the corresponding $\tilde{N}(\tilde{N}-1)/2$ distances. Following Ref.~\cite{Dot05}, we additionally calculate the averaged distances from successively acquired images of the same atoms to reduce the statistical error. 

\begin{figure}[ht]
  \includegraphics[width=0.98\columnwidth]{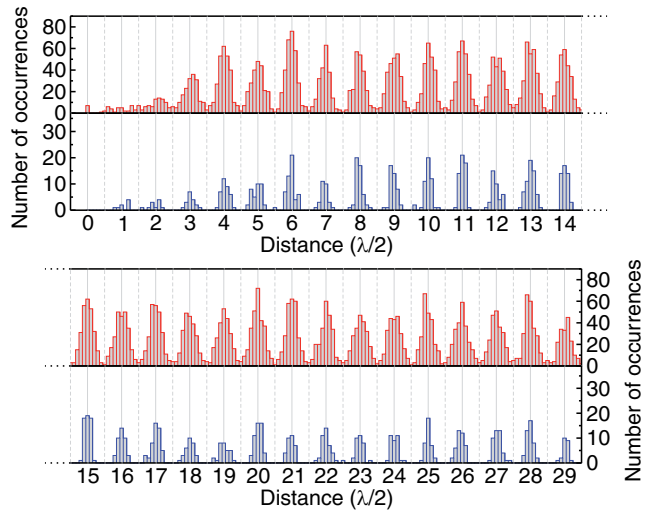}
  \caption{\label{fig:histogram}
  Histogram of determined distances. The upper rows correspond to the distances obtained from a single image; the lower rows to the distances obtained from the averages over 3 successively acquired images of the same atoms.
  }
\end{figure}

In Figure \ref{fig:histogram} the histograms of determined distances from both single images and 3 image averages are shown. Both reveal a periodic structure, perfectly reproducing the periodicity of the optical lattice, as well as a marked decay in the number of occurrences for small separations. We stress that the lattice periodicity does not enter in any way our estimation procedure. Thus, the strict adherence to the periodicity and the well separated histogram peaks provide a striking confirmation for the high precision and reliability of our detection. For each histogram peak, we estimate the reliability $F_{n\in\mathbb{N}}$ of inferring the correct number $n$ of sites separating two atoms, assuming Gaussian distributions of the measured distances around $d_{n}=n\lambda/2$ and fitting a sum of Gaussians centered at $d_{n}$ to the histogram. The reliability $F_{n}$ is then given by the area of the normalized Gaussian centered at $d_{n}$ within $[d_{n}-\lambda/4,d_{n}+\lambda/4]$. For site separations below the diffraction limit ($n<5$) we obtain $F_{\text{1--4}}=68.8\text{--}99.4\,\%$, whereas above the diffraction limit $F_{\text{5--29}}=97.7\text{--}99.8\,\%$. Reducing the statistical error by averaging the distances from $3$ successively acquired images of the same atoms increases the reliability to $F_{\text{1--4}}^{\text(3)}\geq 92.0\,\%$, and $F_{\text{5--29}}^{\text(3)}\geq 99.992\,\%$. This allows us to investigate possible atomic interactions in the nearest-neighbor regime.

The marked decay in the number of occurrences for small separations can partially be traced back to the cases with three or more atoms occupying nearest lattice sites. These cases are challenging from a numerical point of view and cause our algorithm to provide increasingly inaccurate results, which then fail the reliability check. From simulations we deduce that this does not hold for pairs of nearest-neighbors separated by at least two lattice sites from other atoms.

In order to investigate the influence of possible interaction induced effects in the physically interesting regime of neighboring atoms, and to get an unbiased insight in the statistics of atom pair separations, in the following experiment we focus on the distance distribution of isolated atom pairs only. For this purpose, we reduce the mean number of atoms loaded into the MOT to about two atoms using a high field-gradient of $345\,\text{G}/\text{cm}$, which also favors short distances. From single-atom images we infer the interaction-free position distribution, which is related to the overlap of the MOT and the 1D lattice. From two-atom images, we determine the atom separations averaging over 3 successively acquired images of each pair of atoms.

\begin{figure}[ht]
  \includegraphics[width=0.98\columnwidth]{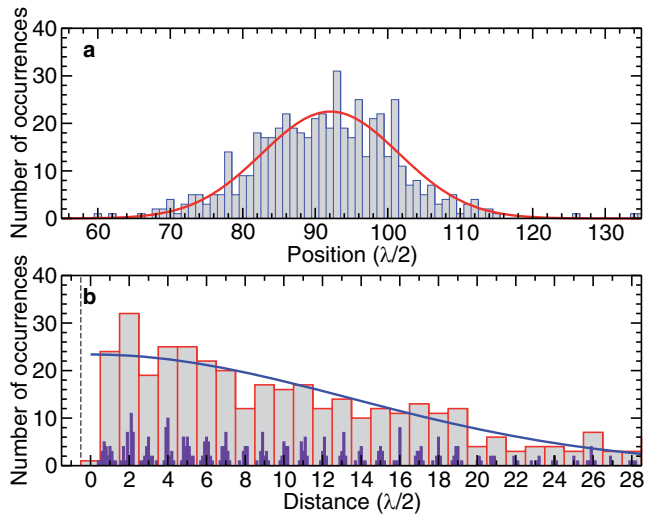}
  \caption{\label{fig:loading}({\bf a}) Histogram of absolute positions of single atoms transferred from the MOT into the 1D lattice. The solid curve shows a Gaussian fit. ({\bf b}) Histogram of distances between two transferred atoms. Solid bars correspond to the measured distances averaged over 3 successively acquired images; shaded bars to the distance distribution in terms of lattice sites. The solid curve shows the  distribution $Q(d)$.
  }
\end{figure}
In Figure \ref{fig:loading}, the histograms of single atom positions and pair separations are shown. The latter reveals a clear gap for two atoms loaded to the same lattice site, reflecting light induced atom losses. The underlying loss mechanism is known \cite{Sesko89,DePue99}: The atoms gain sufficient kinetic energy in an excited molecular potential to leave the trap \cite{footnote1}. For larger separations, starting from nearest-neighbor sites, the distance distribution follows a Gaussian shape. This shape fits well to a modeled distribution which assumes statistically independent positions of both atoms. The number of occurrences $Q$ as a function of distance $d$ is then given by $Q(d)=2 Q_{0}\int_{-\infty}^{\infty}P(x)P(x+d)\text{d}x$,
where $P(x)$ denotes the normalized fitted single atom position distribution (Gaussian with  $\sigma=9.5\times\lambda/2$), see Fig.~\ref{fig:loading}(a), and $Q_{0}$ the total number of analyzed atom pairs. From this fact we conclude that, for our 1D lattice, atoms in neighboring lattice sites do not affect each others' storage time, which is limited by background gas collisions. We stress that for isolated atom pairs with separations below the diffraction limit, our numerical procedure provides an increased reliability of more than $95.0\,\%$, exceeding the number quoted above for clusters of atoms.

Summarizing, we have presented a method to determine the separations of individual neutral atoms in a 1D optical lattice from a fluorescence image at all relevant distances. We have investigated the statistics of atom pair separations and directly observed light induced atom losses for atoms occupying the same lattice site. Our work yields promising perspectives for the implementation and detection of controlled interactions between atoms using, e.g., spin dependent optical potentials \cite{Man03a,Man03b}. Furthermore, since all stages of our numerical processing are extendible to higher dimensions, it points to high resolution imaging of 2D systems, e.g., degenerate lattice gases, or nano dots on substrates in vacuum. For instance, finding a defect in a homogeneously filled Mott-insulator by inverting our method seems conceivable.

\begin{acknowledgments}
We thank D.~D{\"o}ring, A.~H{\"a}rter, F.~Grenz and A.~Rauschenbeutel for technical assistance and A.~Steffen for valuable discussions. This work was supported by the DFG (Research Unit 635) and the EU (SCALA).
\end{acknowledgments}

\vfill

\bibliographystyle{prsty}
\bibliographystyle{physrev}

\begin{thebibliography}{19}
\bibitem{Reliability}
For every excluded data set, we manually investigate the sequence of successive images. For details see \cite{KarskiToBePrep}.
\bibitem{footnote1}
We do not expect measurable modifications of this process in our 1D lattice, because the lattice confinement is too small to induce significant changes of the relative wave function at small inter-nuclear separations, cf.~\cite{DePue99,Sun06}.
\end{thebibliography}

\end{document}